
\input epsf.tex
\input harvmac.tex





\def\rmk#1{\bigskip\noindent{\bf Remarks} }


\def\figin{\epsfcheck\figin}\def\figins{\epsfcheck\figins}
\def\epsfcheck{\ifx\epsfbox\UnDeFiNeD
\message{(NO epsf.tex, FIGURES WILL BE IGNORED)}
\gdef\figin##1{\vskip2in}\gdef\figins##1{\hskip.5in}
\else\message{(FIGURES WILL BE INCLUDED)}%
\gdef\figin##1{##1}\gdef\figins##1{##1}\fi}
\def\DefWarn#1{}
\def\figinsert{\goodbreak\midinsert}
\def\ifig#1#2#3{\DefWarn#1\xdef#1{fig.~\the\figno}
\writedef{#1\leftbracket fig.\noexpand~\the\figno}%
\figinsert\figin{\centerline{#3}}\medskip\centerline{\vbox{\baselineskip12pt
\advance\hsize by -1truein\noindent\footnotefont{\bf Fig.~\the\figno:} #2}}
\bigskip\endinsert\global\advance\figno by1}


\def\IC{\relax{\rm I\kern-.18em C}}

\def\IL{\relax{\rm I\kern-.18em L}}
\def\IH{\relax{\rm I\kern-.18em H}}
\def\IR{\relax{\rm I\kern-.18em R}}
\def\IC{\relax\hbox{$\inbar\kern-.3em{\rm C}$}}
\def\IZ{\relax\ifmmode\mathchoice
{\hbox{\cmss Z\kern-.4em Z}}{\hbox{\cmss Z\kern-.4em Z}}
{\lower.9pt\hbox{\cmsss Z\kern-.4em Z}}
{\lower1.2pt\hbox{\cmsss Z\kern-.4em Z}}\else{\cmss Z\kern-.4em Z}\fi}

\def\CN {{\cal N}}
\def\CR {{\cal R}}

\def\CP {{\cal P }}
\def\CL {{\cal L}}

\def\CS {{\cal S}}

\font\manual=manfnt \def\dbend{\lower3.5pt\hbox{\manual\char127}}

\def\IZ{\relax\ifmmode\mathchoice
{\hbox{\cmss Z\kern-.4em Z}}{\hbox{\cmss Z\kern-.4em Z}}
{\lower.9pt\hbox{\cmsss Z\kern-.4em Z}}
{\lower1.2pt\hbox{\cmsss Z\kern-.4em Z}}\else{\cmss Z\kern-.4em Z}\fi}
\def\half {{1\over 2}}

\def\p{\partial}

\def\CL {{\cal L}}

\def\CN {{\cal N}}

\def\CP {{\cal P }}

\def\CS {{\cal S }}

\def\heis{{\rm Heis}}


\def\IZ{\relax\ifmmode\mathchoice
{\hbox{\cmss Z\kern-.4em Z}}{\hbox{\cmss Z\kern-.4em Z}}
{\lower.9pt\hbox{\cmsss Z\kern-.4em Z}}
{\lower1.2pt\hbox{\cmsss Z\kern-.4em Z}}\else{\cmss Z\kern-.4em
Z}\fi}
\def\IB{\relax{\rm I\kern-.18em B}}
\def\IC{{\relax\hbox{$\inbar\kern-.3em{\rm C}$}}}
\def\ID{\relax{\rm I\kern-.18em D}}
\def\IE{\relax{\rm I\kern-.18em E}}
\def\IF{\relax{\rm I\kern-.18em F}}
\def\IG{\relax\hbox{$\inbar\kern-.3em{\rm G}$}}
\def\IGa{\relax\hbox{${\rm I}\kern-.18em\Gamma$}}
\def\IH{\relax{\rm I\kern-.18em H}}
\def\II{\relax{\rm I\kern-.18em I}}
\def\IK{\relax{\rm I\kern-.18em K}}
\def\IP{\relax{\rm I\kern-.18em P}}

\def\IQ{\relax\hbox{$\inbar\kern-.3em{\rm Q}$}}
\def\IP{\relax{\rm I\kern-.18em P}}

\def\IB{\relax{\rm I\kern-.18em B}}
\def\ID{\relax{\rm I\kern-.18em D}}
\def\IE{\relax{\rm I\kern-.18em E}}
\def\IF{\relax{\rm I\kern-.18em F}}
\def\IG{\relax\hbox{$\inbar\kern-.3em{\rm G}$}}
\def\IGa{\relax\hbox{${\rm I}\kern-.18em\Gamma$}}
\def\IH{\relax{\rm I\kern-.18em H}}
\def\II{\relax{\rm I\kern-.18em I}}
\def\IJ{\relax{\rm I\kern-.18em J}}
\def\IK{\relax{\rm I\kern-.18em K}}
\def\IL{\relax{\rm I\kern-.18em L}}

\def\IN{\relax{\rm I\kern-.18em N}}
\def\IO{\relax{\rm I\kern-.18em O}}
\def\IP{\relax{\rm I\kern-.18em P}}
\def\IQ{\relax\hbox{$\inbar\kern-.3em{\rm Q}$}}
\def\IR{\relax{\rm I\kern-.18em R}}

\def\IW{\relax\hbox{$\inbar\kern-.3em{\rm W}$}}

\def\Im{{\rm Im}}

\def\Hom{{\rm Hom}}
\def\Ext{{\rm Ext}}

\def\inbar{\,\vrule height1.5ex width.4pt depth0pt}

\def\p{\partial}

\font\cmss=cmss10 \font\cmsss=cmss10 at 7pt
\def\IR{\relax{\rm I\kern-.18em R}}


%
\def\inv{^{\raise.15ex\hbox{${\scriptscriptstyle -}$}\kern-.05em 1}}

\def\Dsl{\,\raise.15ex\hbox{/}\mkern-13.5mu D} 
\def\dsl{\raise.15ex\hbox{/}\kern-.57em\partial}





%


\def\unlockat{\catcode`\@=11}
\def\lockat{\catcode`\@=12}

\unlockat

\def\newsec#1{\global\advance\secno by1\message{(\the\secno. #1)}
\global\subsecno=0\global\subsubsecno=0\eqnres@t\noindent
{\bf\the\secno. #1}
\writetoca{{\secsym} {#1}}\par\nobreak\medskip\nobreak}
\global\newcount\subsecno \global\subsecno=0
\def\subsec#1{\global\advance\subsecno
by1\message{(\secsym\the\subsecno. #1)}
\ifnum\lastpenalty>9000\else\bigbreak\fi\global\subsubsecno=0
\noindent{\it\secsym\the\subsecno. #1}
\writetoca{\string\quad {\secsym\the\subsecno.} {#1}}
\par\nobreak\medskip\nobreak}
\global\newcount\subsubsecno \global\subsubsecno=0
\def\subsubsec#1{\global\advance\subsubsecno by1
\message{(\secsym\the\subsecno.\the\subsubsecno. #1)}
\ifnum\lastpenalty>9000\else\bigbreak\fi
\noindent\quad{\secsym\the\subsecno.\the\subsubsecno.}{#1}
\writetoca{\string\qquad{\secsym\the\subsecno.\the\subsubsecno.}{#1}}
\par\nobreak\medskip\nobreak}

\def\subsubseclab#1{\DefWarn#1\xdef
#1{\noexpand\hyperref{}{subsubsection}%
{\secsym\the\subsecno.\the\subsubsecno}%
{\secsym\the\subsecno.\the\subsubsecno}}%
\writedef{#1\leftbracket#1}\wrlabeL{#1=#1}}
\lockat



\def\boxit#1{\vbox{\hrule\hbox{\vrule\kern8pt
\vbox{\hbox{\kern8pt}\hbox{\vbox{#1}}\hbox{\kern8pt}}
\kern8pt\vrule}\hrule}}
\def\mathboxit#1{\vbox{\hrule\hbox{\vrule\kern8pt\vbox{\kern8pt
\hbox{$\displaystyle #1$}\kern8pt}\kern8pt\vrule}\hrule}}


\lref\FreedYA{
  D.~S.~Freed, G.~W.~Moore and G.~Segal,
  ``The uncertainty of fluxes,''
  arXiv:hep-th/0605198.
}
\lref\FreedYC{
  D.~S.~Freed, G.~W.~Moore and G.~Segal,
  ``Heisenberg groups and noncommutative fluxes,''
  arXiv:hep-th/0605200.
}

\lref\Kitaev{A. Kitaev, ``Protected qubit based on a superconducting
current mirror,'' arXiv:cond-mat/0609441}

\lref\Choi{Mahn-Soo Choi, M. Y. Choi, Sung-Ik Lee,
 ``Quantum Phase Transition in Particle-Hole Pair
Transport in Capacitively Coupled Josephson-Junction Chains,''
cond-mat/9802199; Mahn-Soo Choi, ``Capacitively coupled
Josephson-junction chains: straight and slanted coupling,''
 cond-mat/9802237 }
%

\lref\GukovKN{
  S.~Gukov, M.~Rangamani and E.~Witten,
  ``Dibaryons, strings, and branes in AdS orbifold models,''
  JHEP {\bf 9812}, 025 (1998)
  [arXiv:hep-th/9811048].
}

\lref\HirschSmale{ M. Hirsch, ``Immersions of Manifolds,'' Trans.
A.M.S. 93 (1959), 242-276.}

\lref\Kirby{ R. Kirby, {\it The Topology of 4-Manifolds,} Lecture
Notes in Mathematics 1374, Springer Verlag, 1989.}




%
\Title{\vbox{\baselineskip12pt \hbox{ } \hbox{NSF-KITP-07-143} } }
{\vbox{\centerline{Noncommuting Flux Sectors in a Tabletop
Experiment } }} \centerline{Alexei Kitaev$^1$ , Gregory W.
Moore$^{2}$, Kevin Walker$^3$ }

\bigskip

\medskip

\centerline{$^{1}${\it California Institute of Technology}}
\centerline{\it Pasadena, CA 91125, USA}

\smallskip

\centerline{$^{2}${\it Department of Physics, Rutgers University}}
\centerline{\it Piscataway, NJ 08854-8019, USA}

\smallskip

\centerline{$^{3}${\it Microsoft Station Q }} \centerline{\it
University of California, Santa Barbara, 93108 }

\medskip

 \vskip.1in \vskip.1in \centerline{\bf Abstract}
 \medskip
\noindent We show how one can use superconductors and Josephson
junctions to create a laboratory system which can explore the
groundstates of the free electromagnetic field in a 3-manifold with
torsion in its cohomology.

\Date{ June 21, 2007 }


\newsec{ Introduction }

Abelian gauge theories exhibit a curious uncertainty principle
between the topological classes of electric and magnetic flux
sectors. One version of this phenomenon arose in string theory
\GukovKN\ and it has been thoroughly explored in  \FreedYA\FreedYC.
The uncertainty principle even applies to ordinary $3+1$ dimensional
Maxwell theory, and hence it is natural to ask if one could devise
an experiment to demonstrate it. This paper shows that such an
experiment is indeed possible. Moreover, it is related to recent
ideas for designing topologically protected qubits in quantum
computation \Kitaev.

The effect we wish to demonstrate arises when one considers Maxwell
theory in spacetimes of the form $Y \times \IR$ where $Y$ is an
oriented 3-manifold with torsion in its integral cohomology group
$H^2(Y)$. \foot{Except in the appendix, all homology and cohomology
groups in this paper will have coefficients in $\IZ$.} In
particular, the groundstates of the Maxwell theory will form an
irreducible representation of the Heisenberg group
\eqn\heisgroup{ \heis\bigl({\rm Tors} H^2(Y) \times {\rm Tors}
H^2(Y)\bigr) } where the cocycle defining the Heisenberg group is
defined by the link pairing.  This would appear, at first sight, to
be an extremely esoteric observation. Nevertheless, we will show
that the basic phenomenon can  in principle be experimentally
observed in a tabletop experiment using only appropriate arrays of
Josephson junctions.

In trying to devise an experiment that exhibits this phenomenon we
are immediately confronted with a discouraging fact, which was
pointed out to us by M. Freedman:  For any region $\CR\subset \IR^3$
the cohomology groups $H^2(\CR)$ and $H_1(\CR)$ are torsion free.
See the appendix for an explanation. We will show, however, that by
combining superconductors with a new device \Choi\Kitaev\  based on
Josephson junctions one can make identifications on the holonomies
of the gauge field  which (in the limit of low energies and large
capacitance) mimic the identifications needed to define an abstract
3-manifold with torsion in its homology.

The new device may be described as a superconducting current mirror,
herafter referred to as an SCM. It can be realized as a pair of
capacitively coupled Josephson junction chains \Choi , though the
implementation has not yet been achieved experimentally. An ideal
SCM is an electric circuit element with four superconducting leads
whose energy in the absence of a magnetic field is given by
$E=f(\varphi_1-\varphi_2+\varphi_3-\varphi_4)$, where
$\varphi_1,\varphi_2,\varphi_3,\varphi_4$ are the values of the
superconducting phase on the leads and $f$ is a function with a
global minimum at $0$. It has been observed recently in \Kitaev\
that the SCM can be turned into a topologically protected qubit by
connecting the four leads diagonally, which is described by setting
$\varphi_1=\varphi_3$ and $\varphi_2=\varphi_4$. Under  these
circumstances, the energy has two equal minima at
$\varphi_1-\varphi_2=0$ and $\varphi_1-\varphi_2=\pi$. In this paper
we build on the same idea but interpret it differently. While the
above description may be viewed as an ``electrical engineering
approach" where one thinks of an electric circuit in terms of
currents (or superconducting phases), we suggest that the two-fold
degenerate ground state can also be understood as a property of the
electromagnetic field in the free space surrounding the
superconductor.

We now discuss the general principles by which one can map
superconducting circuits to properties of the groundstates of free
Maxwell theory on three-manifolds $Y$. We are aiming to   write an
effective quantum mechanical system for the low energy degrees of
freedom. Consider quantum Maxwell theory in spatial $\IR^3$, but
with a connected region $\CS$ filled with superconductor. This will
be related to Maxwell theory on $\IR^3/\sim$ where $\sim$ identifies
$\CS$ to a single point $\CP$.  The reason is that $\vec E = \vec B
=0$ inside $\CS$, so that inside $\CS$ there is only a flat gauge
field. Suppose, for the moment, that the bosons which condense in
the superconductor have the elementary unit of charge. Then, by flux
quantization, the holonomies\foot{Our conventions for gauge fields
are those of \FreedYA\FreedYC. In particular, $A$ is normalized so
that $F=dA$ locally, and $F$ has {\it integral} periods.}
\eqn\holonm{ \exp\bigl( 2\pi i \int_\gamma A \bigr) }
around homotopically nontrivial cycles $\gamma \subset \CS$ must be
trivial, and hence the gauge field in $\CS$ is trivial. Therefore,
the gauge bundle with connection restricted to $\CS$ is trivial and
$\CS$ can be identified to a point.

Two points raised by the above proposal require further discussion.
  First,   in Nature the condensing bosons -- the Cooper pairs
-- actually have {\it twice} the elementary charge, so the above
argument leaves open the possibility that holonomies around
noncontractible loops in $\CS$ can be $-1$. Let $\CL$ be the line
bundle corresponding to the representation with the elementary
charge. Any superconducting circuit in $\IR^3$ can be described by a
globally defined $U(1)$ connection $A$ on $\CL$.  It is therefore
completely defined by its fieldstrength, which vanishes in the
superconducting region $\CS$. This holds for both $A$ as well as the
connection $2A$ on $\CL \otimes \CL$. However, the latter has
holonomy $=1$ inside $\CS$. Now, we can unambiguously  obtain the
fieldstrength $F(A)$ from that of $F(2A)=2F(A)$, and hence we can
thereby reconstruct the original connection $A$ from $2A$. The
important point is that it is the connection $2A$ which has a nice
description in terms of an effective field theory in the
complementary vacuum region. Of course, field configurations $A$
with holonomy $-1$ around loops in $\CS$ do exist, but these
correspond to higher energy states, and are not important to the low
energy effective field theory.  (In particular, the magnetic energy
of a half flux quantum trapped in a superconducting ring is much
greater than the ground state splitting in an SCM-based qubit.) For
simplicity, in the following we continue to assume that the boson
which condenses in the superconductor has the elementary unit of
charge. By the above remarks we can always map the low energy states
of this hypothetical system to the case where the condensing boson
has twice the elementary charge, as it is in Nature.

The second point is that   the region $\IR^3/\sim$ is not
necessarily a manifold. In formulating Maxwell theory directly on
this space we must use a boundary condition. We assume that at
$\CP$, the point of identification,  the fields are zero.
Alternatively, we can work with 3-manifolds with boundary $Y$ with
superconducting boundary conditions on $\p Y$, i.e., $A$ is trivial
on $\p Y$, in which case the uncertainty principle on topological
sectors is determined by the link pairing (again a perfect pairing):
\eqn\heisgroupbdy{  {\rm Tors} H^2(Y,\p Y ) \times {\rm Tors} H^2(Y)
\to U(1). }
(In general boundary conditions on a free Maxwell field in a
spacetime $M$ are formulated by using the $2$-form $\Omega =
\int_{\p M} \delta A \wedge * \delta F$ to define a symplectic form
on the Hamiltonian reduction of fieldspace. A boundary condition is
a Lagrangian subspace with respect to this form. We choose trivial
connection on $\p Y$ which entails the standard conditions
$E_{\parallel} =0$ and $B_\perp=0$ at the boundary of the
superconductor. If it were possible to condense magnetic monopoles
in nature we could use the electromagnetic dual boundary condition,
and in this case, the construction of qubits would be quite easy. )

\ifig\figone{Layer between two superconductors   . }
{\epsfxsize2.0in\epsfbox{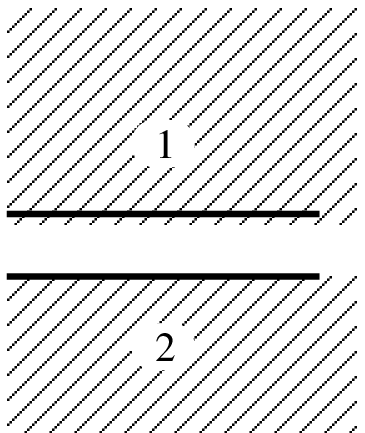}}

Now let us discuss the effective quantum theory. First, in   a
region surrounded by superconductor as in \figone,  the wavefunction
is a function of the gauge-invariant variable
\eqn\defuot{ u_{12} := \varphi_1 - \varphi_2 + 2\pi \int_1^2 A
}
where $\varphi_i$ is the phase of the superconducting condensate and
the contour integral is along a short vertical path from region 1 to
region 2. The variable $u_{12}$ is defined modulo $2\pi \IZ$.

\ifig\figtwo{A SCM with leads $1,2,3,4$. In this paper we treat the
region inside the  dotted  lines as a black box. }
{\epsfxsize3.0in\epsfbox{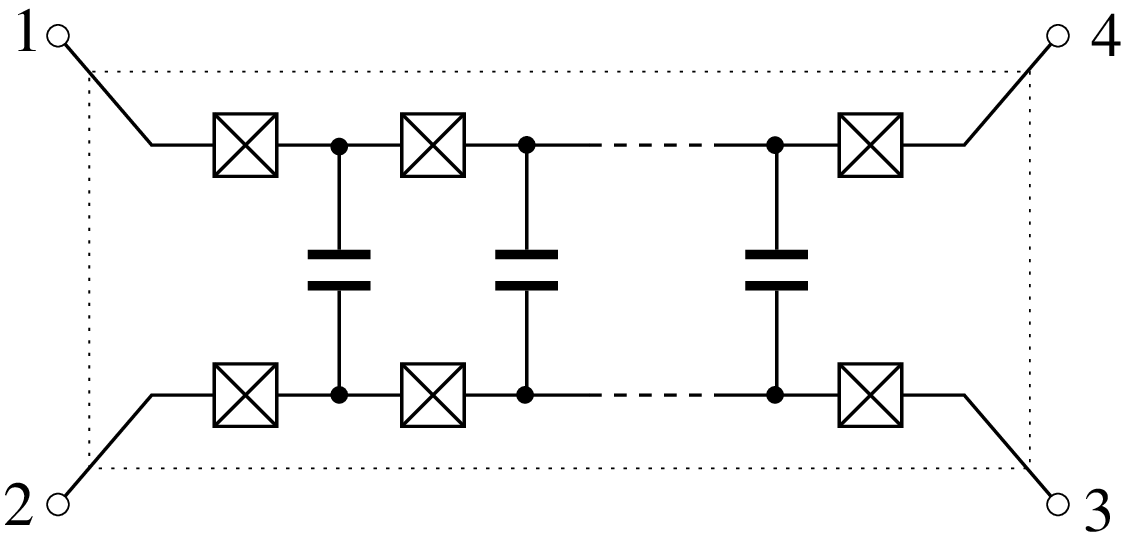}}

Next we consider the superconducting mirror (SCM)   shown in
\figtwo. The SCM adds a term to the Hamiltonian for the low energy
modes given by    $f(u_{14}- u_{23})$ where $f$ is a function of a
periodic variable with a single minimum at $0$. \foot{This is the
origin of the   name ``superconducting mirror.'' It reflects the
fact that the currents $J_i = {\p \over \p u_i} E$, where $E$ is the
energy, are all equal in magnitude. } This is a result of
\Choi\Kitaev. For small devices
 we can consider
the function to be $f(u_{12}+u_{34})$.   Note that in a limit (such
as the semiclassical limit) in which the potential function
dominates the low energy quantum mechanics this imposes the
constraint
\eqn\semcn{ u_{12} + u_{34} = 0 \pmod{2\pi \IZ} }
on the ground state. Note too that we have used
\eqn\fappx{ u_{14}-u_{23} - ( u_{12}+u_{34}) = - \left(\int_1^2 A
+\int_2^3 A + \int_3^4 A + \int_4^1 A \right) }
so, if the magnetic flux through the device is small we can neglect
the right hand side. For definiteness we will sometimes take the
potential term in the Hamiltonian to be:
\eqn\qjaterm{ - J \cos \bigl( u_{12} + u_{34} \bigr) }

\newsec{An Example}

Consider the Klein bottle $K$. We use this to form a twisted
interval bundle
$$I \rightarrow Y\rightarrow K$$
 where the twisting
cancels $w_1(K)$ so that $Y$ is orientable. The boundary $\p Y$ is a
torus - the orientation double cover of $K$. We are going to design
a situation where Maxwell theory is effectively placed in the three
manifold $Y$. \foot{As mentioned above, the theory really lives on
$Y/\p Y$, or on $Y$ with superconducting boundary conditions on $\p
Y$.  } Recall that $\pi_1(K) = \langle a,b\vert aba^{-1}
b=1\rangle$, so the abelianization is $\IZ \times \IZ_2$, and hence
$$
H^2(Y,\p Y) \cong H_1(Y) \cong H_1(K) \cong \IZ\times \IZ_2
$$
has torsion.

\ifig\figthree{A sketch of the manifold $N$. }
{\epsfxsize2.0in\epsfbox{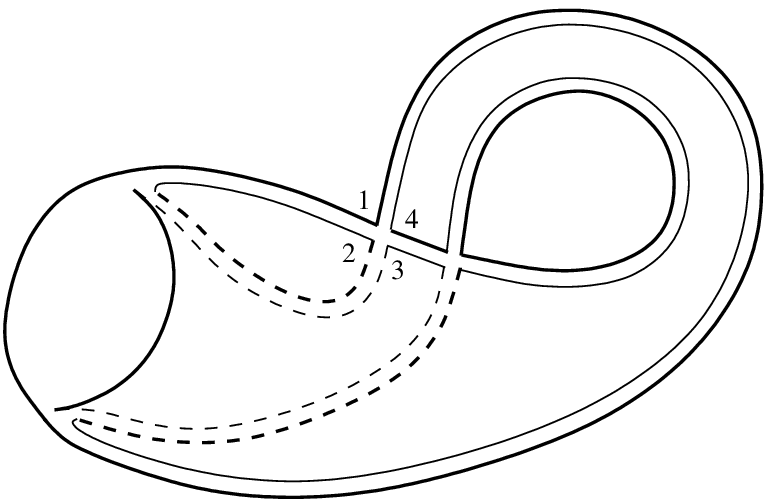}}

Of course, $K$ and $Y$ cannot be embedded into $\IR^3$, but they can
be immersed with only double points.  The double points of the
immersion $i(K)$ trace out a figure $X \times S^1$, where here $X$
is a literal $X$. Consider a thickening of $i(K)$, e.g., the region
occupied by a model of $i(K)$ made with glass. The region occupied
by the glass is   a 3-manifold with boundary which we will denote
$N$. A sketch of $N$ appears in \figthree. The boundary $\p N$ is
the disjoint union of two tori $T^2$, and $N$ itself may be viewed
as  a cobordism from $T^2$ to $T^2$ obtained by cutting a small
solid torus from within a small ball within a larger solid torus.

\ifig\figfour{Region near the double point of  immersed Klein bottle
(dashed line). It has been thickened, and there is superconductor in
the shaded region.   } {\epsfxsize2.0in\epsfbox{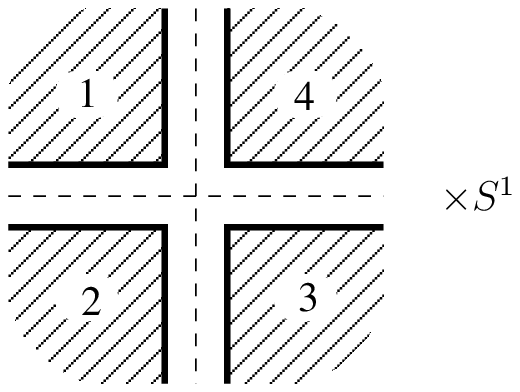}}

 Now imagine that $N$ is
not filled with glass but with vacuum, and that the complementary
region $\IR^3-N$ is filled with
 superconductor, both on the inner region and on the outer region.
Consider the neighborhood of a double point as in \figfour.

A priori there are 4 gauge invariant variables $u_{12}, u_{23},
u_{34}, u_{41}$  satisfying the constraint $\sum u_{i,i+1}=0$.
However, in the complement of $N$   a point in region 1 is
continuously connected  to a point in region 3, and similarly a
point in region 2 to a point in region 4. Therefore, if the gauge
field is zero there is only one independent variable, say
$u=u_{12}$.   The effective Hamiltonian is given by
$$
H = {Q^2\over 2C}
$$
where $Q=e^* \CN$  is the charge for $\CN$ Cooper pairs of charge
$e^*$ and $C$ is an effective capacitance. The superconducting state
is not a state of definite $\CN$ but is rather described by a
wavefunction of the conjugate variable $u$ so that $\CN = -i   {\p
\over \p u}$ so that
\eqn\hamilt{ H = - {(e^*)^2 \over 2 C} \bigl({\p \over \p u}
\bigr)^2 }
Since $u \sim u + 2\pi \IZ$, the   gauge invariant configuration
space is a circle. There is no potential function, and hence there
is a unique normalizable ground state $\Psi_{grnd}(u) = constant$.

Now, let us consider the effect of inserting an SCM at a fixed angle
$\theta\in S^1$ in the set of double points $X \times S^1$ of
$i(K)$. According to \Kitaev\ this   adds a term \qjaterm\ to the
Hamiltonian. Because of the topology $u_{12}= u_{34}= u$. In terms
of the circuit in \figtwo\ we would be connecting leads 1 to 3 and
leads 2 to  4. Thus the effective Hamiltonian is
\eqn\hamiltp{ H = - {(e^* )^2 \over 2 C} \biggl({\p \over \p u}
\biggr)^2 - J \cos(2u) }
For $JC\gg 1$, $J>0$ the groundstates are well-approximated by
states localized near $u=0,\pi$, denoted $\vert 0\rangle $ and
$\vert \pi \rangle$. Thus the space of groundstates is effectively
2-dimensional.\foot{We are neglecting some physical effects, e.g.
electron tunneling through the vacuum,   which only give
exponentially small contributions to the ground state splitting. }

As an aside, we note that the   Schr\"odinger equation for this
potential is the well-known Mathieu equation and can be ``solved
exactly.'' The groundstate is of course unique and, for $JC\gg 1$,
closely approximated by ${1\over \sqrt{2}}(\vert 0\rangle +\vert \pi
\rangle)$ . However, in the limit
\eqn\limte{ q:= - {CJ\over (e^* )^2} \rightarrow -\infty  }
there are two low-lying states with energy eigenvalues
\eqn\enspacing{ E_1- E_0   \sim J \vert q
\vert^{-1/4}e^{-4\sqrt{\vert q\vert}} }
(we drop a numerical constant). This confirms and quantifies our
expectation that, to exponential accuracy there is a two-dimensional
space of degenerate groundstates.

We claim that  this 2-dimensional space of approximate groundstates
naturally forms the irreducible representation of the Heisenberg
group $\heis(\IZ_2 \times \IZ_2)$.  The Heisenberg operators
corresponding to the clock and shift operators are
\eqn\clockshift{ P= \pmatrix{1&0 \cr 0 & -1 \cr} \qquad Q =
\pmatrix{0 & 1 \cr 1 & 0 \cr}. }

According to the principles discussed above, effectively the theory
has been put on the space $Y/\p Y$. In the Maxwell theory picture
$P,Q$ correspond to measuring magnetic and electric fluxes, that is,
the magnetic and electric first Chern class $c_1$, respectively. Let
us discuss the physical implementation of these operations in the
corresponding superconducting circuit, following \Kitaev. In the
analogous SCM as in \figtwo\ the leads $1$ and $3$ are connected and
the leads $2$ and $4$ are connected. The operation of $P$
corresponds to inserting a device between leads $1$ and $4$ to
measure the phase $\exp(iu)$. The operator $Q$ is trickier.

\ifig\figQ{A setup for the realization of the shift operator $Q$ or
the measurement in the eigenbasis of $Q$.}
{\epsfxsize1.5in\epsfbox{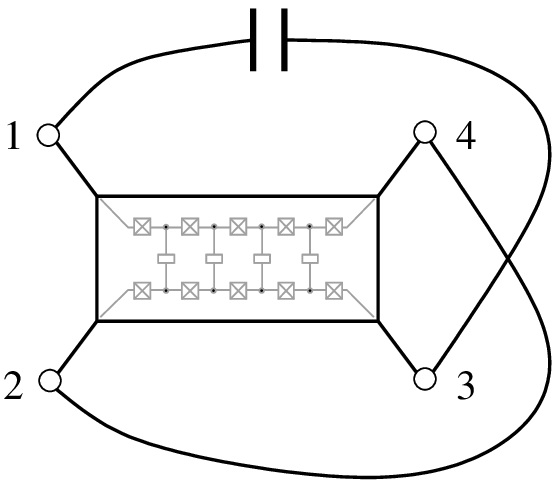}}

The operator $Q$ as a unitary transformation of the quantum state
corresponds to the adiabatic change of $\varphi_1-\varphi_3$ from
$0$ to $2\pi$ or $-2\pi$.  To realize this transformation, one needs
to insert a capacitor on the wire connecting nodes $1$ and $3$,
breaking the identification $u_{13}=0$ (see \figQ). Then the
(classical) groundstate equation becomes
\eqn\gndst{ u_{13} + 2 u_{32} =0 \pmod{2\pi} }
If we increase $u_{13}$ from $0$ to $2\pi$ adiabatically, then
$u_{32}$ shifts from $0$ to $-\pi$. Therefore, after reconnecting
the terminals, $u_{12}$ has shifted from $0$ to $-\pi \cong +\pi$,
and we have implemented the shift operator $Q$.

 The above argument also shows that $Q=e^{2\pi i n}$, where
$n=-i\,\p/\p u_{13}$ is the operator of electric charge on either
capacitor plate. By measuring this charge, one can perform a
measurement in the eigenbasis of $Q$. The charge is related to the
electric field in the capacitor and is therefore observable, though
a practical implementation of the measurement might ultimately use a
different principle. If the SCM is in state ${1\over \sqrt{2}}
(|u\cong 0 \rangle+|u\cong \pi \rangle)$, then the charge takes on
integer values, otherwise the charge is half-integer.

\newsec{Generalization to other Heisenberg groups }

\subsec{An array of SCM's}

\ifig\figsixb{Connecting $n=3$ SCM's as described in the text. For
the implementation of the corresponding shift operator, a capacitor
may be inserted at point C.}
{\epsfxsize2.0in\epsfbox{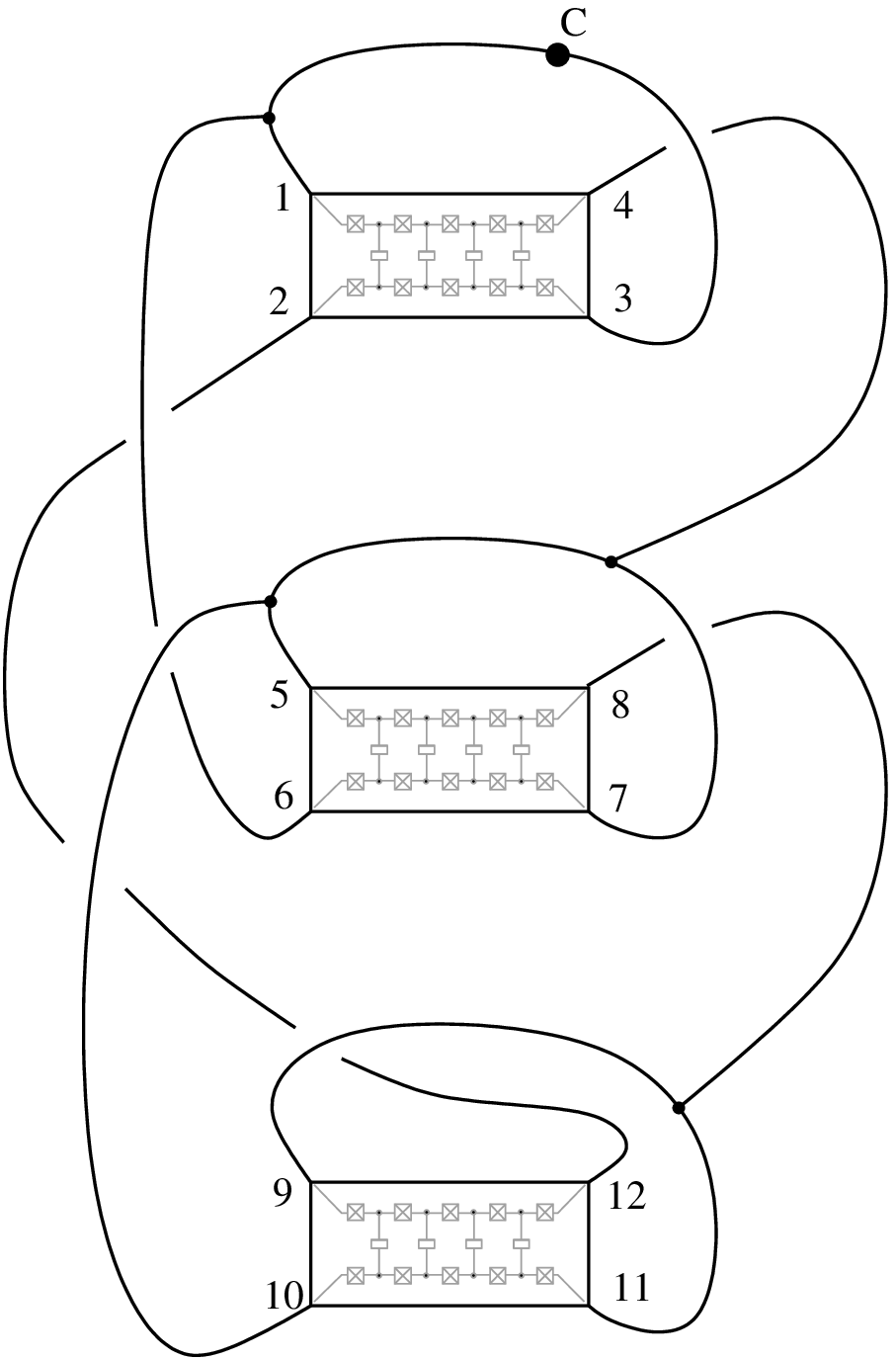}}

Let us consider a system of $n$ SCM's
which will be, roughly speaking, connected in series. The effective
Hamiltonian in the $JC\gg 1$ limit sets
\eqn\zerone{ \eqalign{
 \varphi_1 - \varphi_2 + \varphi_3 -\varphi_4 & = 0 \cr
 \varphi_5 - \varphi_6 + \varphi_7 -\varphi_8 & = 0 \cr
 \varphi_9 - \varphi_{10} + \varphi_{11} -\varphi_{12}  &  = 0 \cr
 \cdots &  \cdots \cr
 \varphi_{4n-3} - \varphi_{4n-2} + \varphi_{4n-1} -\varphi_{4n}  &  = 0
 \cr}
 }
Now connect the wires so that $\varphi_1 = \varphi_3, \varphi_5 =
\varphi_7 , \dots, \varphi_{4j+1}=\varphi_{4j+3}, \dots,
\varphi_{4n-3}=\varphi_{4n-1}$. We further connect wires so that
\eqn\furtheridents{ \eqalign{ &\qquad\  \varphi_1 = \varphi_6  \cr
&\varphi_4 = \varphi_5 =\varphi_{10} \cr &\varphi_8 =
\varphi_9=\varphi_{14} \cr  & \cdots \cr &\varphi_{4n-4} =
\varphi_{4n-3} \cr} }
that is, $\varphi_{4j}=\varphi_{4j+1}=\varphi_{4j+6}$ for $1\leq
j\leq n-2$,  and finally connect $\varphi_{4n} = \varphi_2$. See
\figsixb\ for the case $n=3$.

With the above connections the   groundstate equations become:
\eqn\aennes{ \pmatrix{2 & -1 & 0 & \cdots & \cdots & 0 \cr -1& 2& -1
& \cdots & \cdots & 0 \cr 0 & -1 & 2 & \cdots  & \cdots & 0  \cr
\cdots & \cdots & \cdots & \cdots & \cdots & \cdots \cr \cdots &
\cdots & \cdots & \cdots & \cdots & \cdots \cr \cdots & \cdots &
\cdots & \cdots & \cdots & \cdots \cr 0 & \cdots & \cdots & 2 & -1 &
0 \cr 0 & \cdots & \cdots  & -1&   2 & -1\cr 0 & \cdots & \cdots  &
0 & -1 & 2\cr}\pmatrix{u_1\cr u_2\cr \cdot\cr \cdot \cr \cdot \cr
\cdot\cr \cdot \cr \cdot \cr u_n\cr} =0 }
where
\eqn\uvarsnn{
 \eqalign{ u_1 & = \varphi_1 - \varphi_2 \cr u_2 & =
\varphi_5 - \varphi_2\cr u_3 & = \varphi_9 - \varphi_2\cr \cdots &
\cdots \cr u_n & = \varphi_{4n-3} - \varphi_2 \cr} }
 The solution is
$u_j = j u_1$ and
\eqn\nfold{ (n+1) u_1 =0 \pmod{2\pi\IZ} }

Incidentally, a model  Hamiltonian analogous to \hamilt\  is
\eqn\multiham{ H = - {(e^*)^2\over 2C} \sum_i \bigl({\p \over \p
u_i}\bigr)^2 -\half J \sum_{\alpha>0, {\rm simple}}(e^{i \alpha\cdot
\phi} + e^{-i \alpha\cdot \phi}  ) }
where $\phi = \sum \alpha_i u_i$ and $\alpha_i$ are the simple roots
of $A_{n}$, and $u_i \sim u_i + 2\pi$.  This system is very closely
related to the exactly soluble Toda system. \foot{but according to
Sergei Lukyanov, our system is is not  integrable.} However, as in
the $n=1$ case, for $(CJ)>> (e^*\hbar )^2$ there are - to
exponentially good accuracy -- $(n+1)$ degenerate groundstates. One
natural basis is given by $\vert r\rangle:=\vert u_1 = {r\over n+1}
2\pi \rangle$, $r=0,\dots, n$.

We now claim that this set of groundstates can be regarded as the
irreducible representation of $\heis(\IZ_{n+1}\times \IZ_{n+1} ) $,
generalizing the example we studied previously.  To justify this
claim we need to explain how to implement the standard clock and
shift operators defining the irreducible representation of ${\rm
Heis}(\IZ_{n+1}\times \IZ_{n+1})$.  The clock operator is
implemented by measuring the phase, say, of $u_1 = \varphi_{12}$ --
something which is experimentally quite feasible. The shift operator
is performed in a way analogous to the case $n=1$ (cf.\ \figQ).
First, we place a capacitor at point C in \figsixb\  to break the
relation $\varphi_1=\varphi_3$. Next we adiabatically change the
phase on the capacitor. The classical groundstate equations are
modified so that the first equation in \aennes\ is changed to
\eqn\newclsgr{ 2u_1-u_2 = \varphi_{13} }
while the remaining equations in \aennes\ are unchanged. These
equations imply $u_i={n+1-i\over n+1}\varphi_{13}$ and in particular
\eqn\newshft{ (n+1) u_1 - n \varphi_{13} =0 }
so, increasing $\varphi_{13}$ from $0$ to $2\pi$ adiabatically
results in a phase shift of $u_1$ by $ {n\over n+1} 2\pi$. After
reconnecting leads $1$ and $3$, $u_1$ has shifted by $ {n\over n+1}
2\pi \cong -{1\over n+1}2\pi$. We have thus implemented the (inverse
of the) shift operator.

\subsec{A corresponding 3-dimensional space}

In this section we construct a three-dimensional space $Y$ which
perfectly reproduces the identifications made in the above array of
SCM's.

\ifig\figseven{The basic bordism $\kappa$ which can be concatenated
in series. } {\epsfxsize2.0in\epsfbox{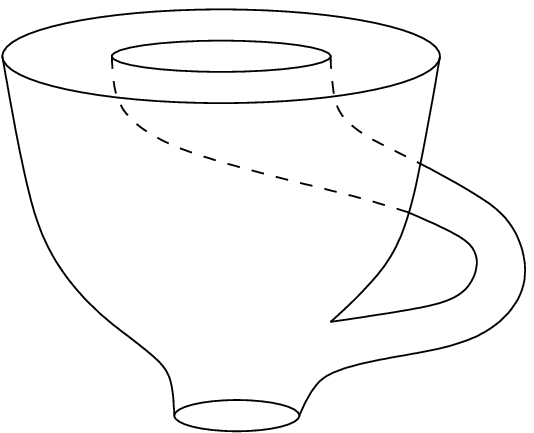}}

Consider first an immersed bordism in $\IR^2 \times I$ from two
concentric circles to a single circle. Considering time evolution
from the top to the bottom, at    the top there are two concentric
tubes. Then the inner tube passes through the outer tube as in the
standard immersion of the Klein bottle. Beyond this point a slice
$\IR^2 \times \{ t\}$ intersects the surface in two nonintersecting
nonconcentric circles. We now adjoin the standard 3-punctured sphere
to give a bordism to a single circle. This bordism - which might be
called the ``Klein jug''  is shown in \figseven\ and will be denoted
by $\kappa$.  If we cap off the bottom circle in $\kappa$ and let
the two top concentric circles merge then we get the standard
immersion of the Klein bottle.

\ifig\figsixa{The case of $n=3$. This corresponds to the circuit in
\figsixb. } {\epsfxsize2.0in\epsfbox{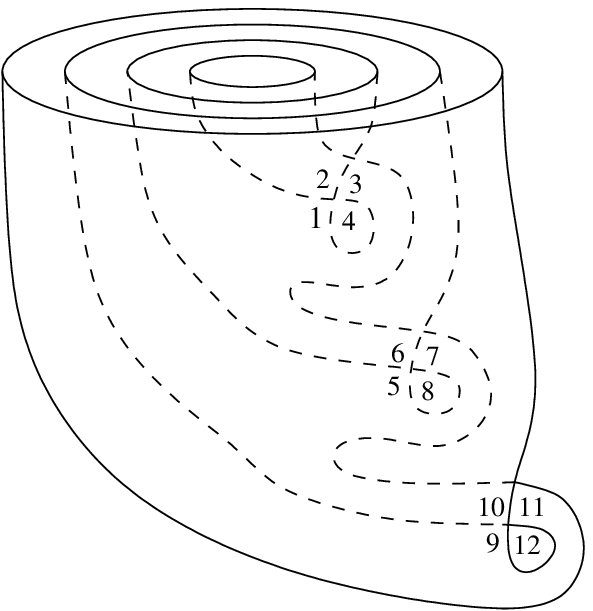}}

To produce the space in the $n>1$ case we consider successive
applications of the bordism $\kappa$. At the top we have $n+1$
concentric circles. We apply $\kappa$ to the two innermost circles
to obtain $n$ concentric circles and continue until there is a
single circle at the bottom. We cap off the bottom circle and fuse
the top $n+1$ circles into a single circle. The case $n=3$ is
illustrated in \figsixa.

Labeling the regions $1,\dots, 4n$ we find that the topology of this
space precisely  implements the identifications made above.

\ifig\Kevinfigure{Cell complex for the ideal space $Y$ for the case
$n=3$. } {\epsfxsize2.0in\epsfbox{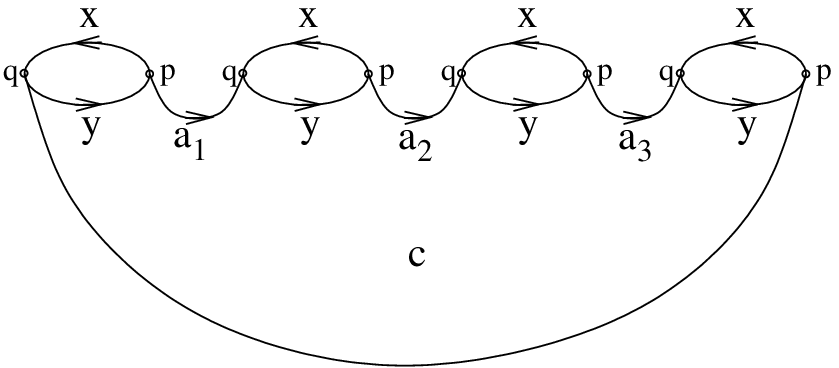}}

The ideal space $Y$ which has been immersed as above can be
described as a 3-dimensional neighborhood of a 2-complex $L$ (i.e.
there is a deformation retraction of $Y$ onto $L$). Therefore
$H^2(Y) \cong H^2(L)$. The 2-complex $L$ can be thought of as an
$n+1$-punctured sphere with each of its boundary components
identified in an orientation preserving fashion with a single
circle.  Accordingly, $L$ has a cell decomposition consisting of two
$0$-cells $p$ and $q$, $n+2$ $1$-cells $x$, $y$, $a_1, \ldots a_n$,
and a single $2$-cell $c$. (See \Kevinfigure.)  The attaching maps
for the $1$- and $2$-cells are given by

\eqn\attachmaps{ \eqalign{
  \partial a_i & = q-p\cr
  \partial x & = q-p\cr
  \partial y & = p-q\cr
  \partial c & = y a_1 y a_2 \cdots a_n y x a_n^{-1} x \cdots x a_1^{-1}
  x\cr}}
It follows that the 1-cycles are freely generated by $x+y, a_1+y,
\ldots a_n+y$, and the $1$-boundaries are generated by $(n+1)(x+y)$.
So $H_1(L)$ has rank $n$ and  ${\rm Tor}(H_1(L)) \cong \IZ_{n+1}$.
Also, there are no 2-cycles, so $H_2(L) \cong 0$.  It now follows
from the universal coefficient theorem  that $H^2(L) \cong
\IZ_{n+1}$. To see that the concatenated Klein jugs are indeed an
immersion of $Y$, pull out the handles of all the $n$ jugs to obtain
a sphere with $(n+1)$ holes with its boundary circles identified.

\ifig\KevinfigureII{The only kinds of singular points of immersions
we need. } {\epsfxsize2.0in\epsfbox{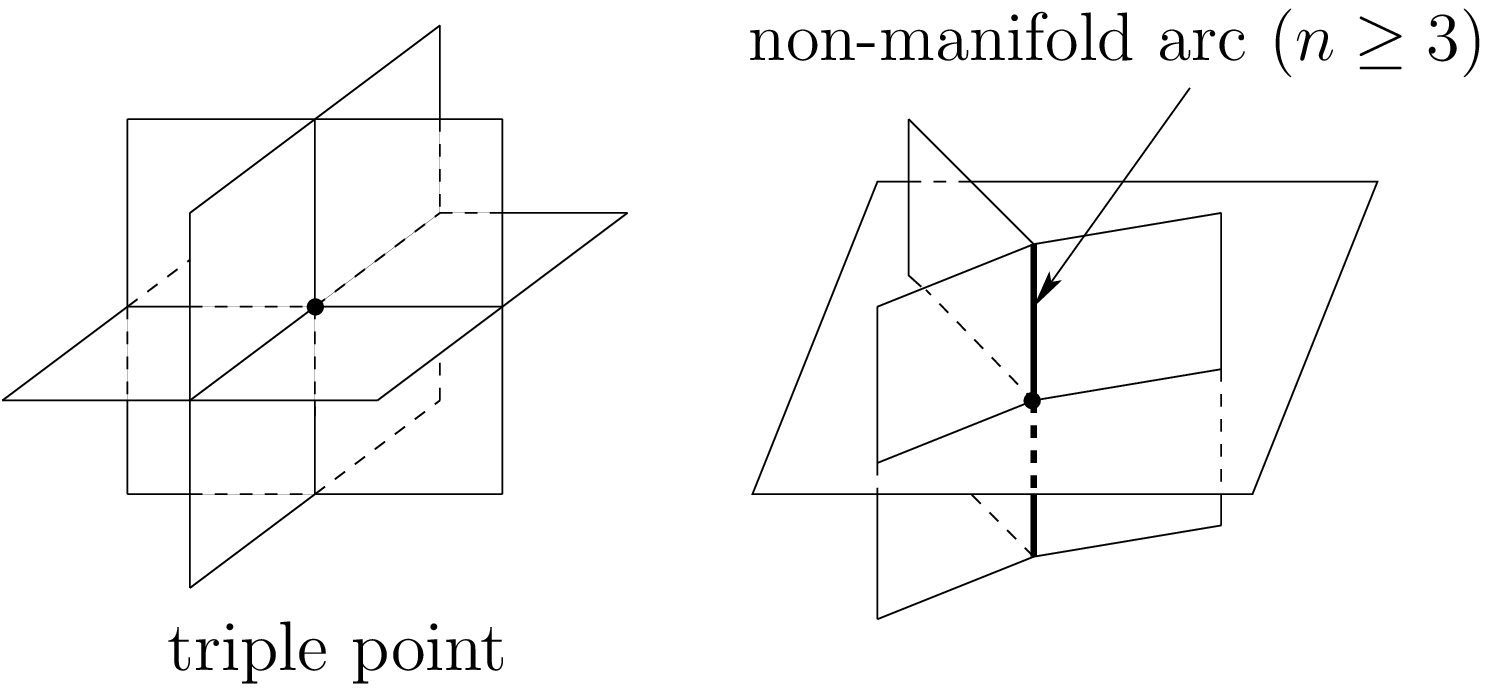}}

 In fact, any
oriented, connected 3-manifold with non-empty boundary can be
immersed in $\IR^3$ as a neighborhood of a $2$-complex with only
double and triple points.  Let $M$ be such a $3$-manifold and let
$f:M \to \IR^3$ be a constant map (all of $M$ sent to a single point
in $\IR^3$).  Since the tangent bundle of $M$ is trivializable
\Kirby, $f$ can be covered by a rank $3$ bundle map $f' : TM \to
T\IR^3$.  It now follows from the Smale-Hirsch immersion theorem
\HirschSmale\Kirby\  that $f'$ can be deformed through rank $3$
bundle maps to the tangent map of an immersion $g: M \to \IR^3$.
(Note that here we use the fact that $M$ has non-empty boundary,
since otherwise the hypotheses of the immersion theorem would not be
satisfied.) Choose a deformation retraction of $M$ onto a
$2$-complex $L \subset M$. (This is equivalent to choosing a handle
decomposition of $M$ which contains no $3$-handles.)  Deform $g$ so
that its restriction to $L$ is a general position map with only
double and triple points.\foot{Once we have immersed the 3-manifold,
we know that any small neighborhood in $L$ is embedded -- there are
no local singularities. Standard results on transversality allow us
to assume, after a small perturbation, that the dimension of the
intersection of the $i$-skeleton of $L$ and the $j$-skeleton of $L$
has dimension $i+j-3$. Similarly, the dimension of the triple
intersection of the $i$, $j$ and $k$-skeleta of $L$ has dimension
$i+j+k-6$. The only possibilities are (a) $2+2-3=1$, $2+1-3=0$ (we
can assume that the $1$-skeleton of $L$ coincides with the
non-manifold points of $L$), and (c) $2+2+2-6=0$. } See
\KevinfigureII.  Using the deformation retraction, we can further
deform $g$ so that $g(M)$ is a small neighborhood of $g(L)$.

We have explained above how to incorporate double points. Triple points do not
require any special treatment: it is sufficient to include one SCM for each
arc of double points.

\bigskip
\noindent{\bf Acknowledgements:}

We would like to thank Michael Freedman for an important
conversation.  G.M. is supported in part by DOE grant
DE-FG02-96ER4094.  A.K. is supported in part by ARO under Grants
No.\ W911NF-04-1-0236 and W911NF-05-1-0294, and by NSF under Grant
No.\ PHY-0456720. This work was initiated at the KITP in December
2005. G.M. thanks the KITP for hospitality. KITP research is
supported in part by the National Science Foundation under Grant No.
PHY99-07949.

\appendix{A}{Properties of torsion cohomology.}

In this appendix we construct the perfect pairing
\eqn\perfpairing{\omega :\
{\rm Tors} H^2(Y;\IZ) \times {\rm Tors} H^2(Y,\p Y;\IZ)
\to\IR/\IZ. }
for any orientable 3-manifold $Y$ with boundary. We also prove the following

{\bf Theorem}: If $Y$ is an embedded 3-manifold in $\IR^3$ then
\eqn\thmstat{ {\rm Tors}(H^2(Y;\IZ)) = {\rm Tors}(H^2(Y;\p
Y,\IZ))=0.}

In fact, such an embedded manifold  cannot even have a finite
fundamental group, as explained to us by Michael Freedman.

Let $Y$ be an orientable 3-manifold; without loss of generality
we may assume that $Y$ is connected. Recall from the coefficient
sequence
\eqn\bockone{ \cdots \rightarrow H^1(Y;\IR) \rightarrow
H^1(Y;\IR/\IZ) ~ {\buildrel \beta \over  \rightarrow} ~ H^2(Y;\IZ)
\rightarrow H^2(Y;\IR) \rightarrow \cdots }
where $\beta$ is the Bockstein map that we can identify ${\rm
Tors}(H^2(Y;\IZ))$ with the image of $\beta$. (Indeed, $\Im\beta$ is
finite since $\beta$ is a continuous map from a compact group to a
discrete group.  Conversely, any torsion element in $H^2(Y;\IZ)$
vanishes if we extend the coefficient group to $\IR$.) The same is
true for the relative cohomology $H^2(Y,\p Y;\IZ)$.

Next, consider the cup-products
\eqn\perfectone{ H^1(Y;\IR/\IZ) \times H^2(Y,\p Y;\IZ) \rightarrow
H^3(Y,\p Y;\IR/\IZ) \cong \IR/\IZ }
\eqn\perfecttwo{ H^2(Y;\IZ) \times H^1(Y,\p Y;\IR/\IZ) \rightarrow
H^3(Y,\p Y;\IR/\IZ) \cong \IR/\IZ }
By Poincar\'e duality, $H^{2}(Y,\p Y;\IZ)\cong H_{1}(Y)$ and
$H^{2}(Y;\IZ)\cong H_{1}(Y,\p Y)$. The above maps are just the
standard pairings between (the absolute or relative) homology and
cohomology in dimension $1$. They can also be interpreted as the
homomorphisms $H^1(\cdots;\IR/\IZ)\to \Hom(H_{1}(\cdots),\IR/\IZ)$
from the universal coefficient sequence, which are actually
isomorphisms since $\Ext^{1}(G,\IR/\IZ)=0$ for any $G$. Thus we have
perfect pairings between the compact group $H^1(Y;\IR/\IZ)$ and the
discrete group $H^{2}(Y,\p Y;\IZ)$, and also between $H^1(Y,\p
Y;\IR/\IZ)$ and $H^{2}(Y;\IZ)$ (Pontryagin duality).

The perfect pairing \perfpairing\ can be obtained from the
cup-products \perfectone\ and \perfecttwo\ if we set
\eqn\perfpdef{ \omega(\beta h,\beta h') \;{\buildrel{\rm def} \over =}\;
h\cup \beta h' = \beta h \cup h',}
where $h\in H^1(Y;\IR/\IZ)$ and $h'\in H^1(Y,\p Y;\IR/\IZ)$. Both
defining expressions are necessary to make sure that $\omega$
depends only on $\beta h$, $\beta h'$ rather than $h$, $h'$.  But we
need to demonstrate that the two definitions are equivalent. To this
end, let us represent the cohomology classes $h$, $h'$ by simplicial
cochains $c$, $c'$ with real coefficients. Then $\beta h$ and $\beta
h'$ are represented by {\it integral} cocycles $dc$ and $dc'$,
respectively. Passing to cohomology classes in the equation
$$
d(c \cup c') = dc \cup c' - c \cup dc'
$$
and taking the quotient over $\IZ$, we get
$$
0 = \beta h \cup h' - h \cup \beta h'.
$$

Now, if $Y$ is an embedded $3$-fold in $S^3$ then its complement $X$
is also an embedded $3$-fold and $X\cup Y = S^3$, while $X \cap Y =
\p X = \p Y$. From the exact sequence for the pair $(S^3,X)$ we
learn
\eqn\pairseqa{ H^1(X;G) \cong H^2(S^3,X;G) \cong H^2(Y,\p Y;G) }
for any coefficient group $G$. Also we have
\eqn\pairseqa{ H^0(X;G)/G \cong H^1(S^3,X;G) \cong H^1(Y,\p Y;G) }
where the quotient by $G$ is by the $0$-cycles which are the same
constant on all the components of $X$.

Now we consider the commutative square
\eqn\comsquare{ \matrix{ H^0(X;\IR/\IZ)/(\IR/\IZ) & \rightarrow &
H^1(X;\IZ) \cr \cong\downarrow  &  & \downarrow\cong \cr H^1(Y,\p
Y;\IR/\IZ) & \rightarrow & H^2(Y,\p Y;\IZ) \cr} }
However, the image of the Bockstein map $H^0(X;\IR/\IZ)/(\IR/\IZ)
\rightarrow  H^1(X;\IZ)$ must be zero since the domain is the direct
product of copies of $\IR/\IZ$ and the image is a discrete
group. Thus ${\rm Tors}(H^2(Y,\p Y; \IZ))$ is zero and by the
perfect pairing ${\rm Tors}(H^2(Y;\IZ))$ is also zero.

 \listrefs

\bye